\documentclass[twocolumn,aps,superscriptaddress,showpacs,nofootinbib,floatfix]{revtex4}
\usepackage{bm,feynmf}
\usepackage{epsfig,bm,feynmf}
\usepackage{graphics}
\usepackage{graphicx}
\usepackage{amsmath}
\usepackage{dcolumn}
\usepackage{xcolor}
\usepackage[normalem]{ulem}  % \sout{old text} for strikeout
\usepackage{tikz-feynman,contour}
\tikzfeynmanset{compat=1.0.0}
\renewcommand\sout{\bgroup \color{red} \ULdepth=-.5ex \ULset}

\begin{document}

%%%%%%%%%%%%%%%%%%%%% Title %%%%%%%%%%%%%%%%%%%%%%

\title{Prerequisites for heavy quark coalescence in heavy-ion collisions}

%%%%%%%%%%%%%%%%%%%% Authors %%%%%%%%%%%%%%%%%%%%%

\author{Taesoo Song}\email{t.song@gsi.de}
\affiliation{GSI Helmholtzzentrum f\"{u}r Schwerionenforschung GmbH, Planckstrasse 1, 64291 Darmstadt, Germany}

\author{Gabriele Coci}\email{g.coci@gsi.de}
\affiliation{GSI Helmholtzzentrum f\"{u}r Schwerionenforschung GmbH, Planckstrasse 1, 64291 Darmstadt, Germany}

%\author{Joerg Aichelin}\email{aichelin@subatech.in2p3.fr}
%\affiliation{SUBATECH UMR 6457 (IMT Atlantique, Universit\'{e} de Nantes, IN2P3/CNRS), 4 Rue Alfred Kastler, F-44307 Nantes, France}
%\affiliation{Frankfurt Institute for Advanced Studies, Ruth-Moufang-Strasse 1, 60438 Frankfurt am Main, Germany}

%\author{Elena Bratkovskaya}\email{E.Bratkovskaya@gsi.de}
%\affiliation{GSI Helmholtzzentrum f\"{u}r Schwerionenforschung GmbH, Planckstrasse 1, 64291 Darmstadt, Germany}
%\affiliation{Institute for Theoretical Physics, Johann Wolfgang Goethe Universit\"{a}t, Frankfurt am Main, Germany}

%%%%%%%%%%%%%%%%%%%% Abstract %%%%%%%%%%%%%%%%%%%%%

\begin{abstract}
The coalescence model assumes instant formation of a bound state from unbound particles based on the overlapping of two states in spatial and momentum spaces and quantum numbers.
Therefore, applied to the hadronization of partons, it provides a snapshot of a Quark-Gluon Plasma (QGP) just before hadronization.
We use the coalescence model for the formation of the ground state of open heavy flavor and the statistical model for heavier states. 
Assuming that all heavy flavors in thermal equilibrium hadronize through the coalescence, 
we find that the QGP just before hadronization is not composed of completely randomized partons but must have strong correlations in color charges 
as well as in momentum and/or coordinate spaces between heavy quark and light (anti-)quark.
\end{abstract}

%\pacs{25.75.Nq, 25.75.Ld}
%\keywords{}

\maketitle

\section{Introduction}

Heavy flavor is one of promising probes to reveal the properties of a hot 
and dense nuclear matter produced in relativistic heavy-ion collisions~\cite{Uphoff:2012gb,He:2014cla,Cao:2013ita,Gossiaux:2010yx,Das:2015ana,Song:2015sfa,Song:2015ykw,Plumari:2017ntm}.
One advantage as the probe particle is that the production of heavy quark 
is reliably described by perturbative Quantum Chromodynamics.
On the other hand, the hadronization of heavy quark into open heavy flavor is a non-perturbative process which requires a phenomenological model.
If the momentum of heavy quark is much larger than heavy quark 
mass, its hadronization can be modelled with a simple fragmentation function~\cite{Peterson:1982ak}.
However, it has been shown that at low transverse momentum the hadronization by only fragmentation is not able to describe the build-up of significant elliptic flow which is measured at RHIC and LHC for light- and 
especially for heavy-flavor hadrons \cite{Adamczyk:2017xur,Acharya:2017qps}. Therefore, one has to rely on a different hadronization approach.

Last two decades the parton coalescence model has successfully described experimental data on particle spectra, elliptic flows, and baryon-to-meson ratios at low and intermediate transverse momenta in relativistic heavy-ion collisions~\cite{Baltz:1995tv,Greco:2003xt,Fries:2008hs}. 
It also has been applied to both open and hidden heavy flavors~\cite{Greco:2003vf,Song:2016lfv,Song:2017phm} as well as to exotic hadrons~\cite{Cho:2017dcy}.
The model assumes that the transition from unbound partons to a bound hadron such as meson and (anti-)baryon takes place instantly and, as a result, the coalescence probability is proportional to Wigner function which describes the spatial and momentum distribution of quarks inside the hadron and therefore depends on relative momentum and distance of 
partons~\cite{Scheibl:1998tk,Song:2016lfv}.

In phenomenological models for the production of heavy flavors in heavy ion collisions, heavy quarks are dominantly hadronized through the coalescence at low transverse momentum while they are hadronized through fragmentation at large transverse momentum~\cite{Uphoff:2012gb,He:2014cla,Cao:2013ita,Gossiaux:2010yx,Das:2015ana,Song:2015sfa,Song:2015ykw,Plumari:2017ntm}.
This can be explained by the fact that a heavy quark with a large transverse momentum is far away from the  light (anti-)quarks in momentum space and, 
as result, the coalescence probability is very small~\cite{Song:2015sfa,Song:2015ykw,Plumari:2017ntm}.
On the other hand, a heavy quark with a small transverse momentum which is surrounded by light (anti-)quarks is close to their phase space and so it has a large probability for coalescence.
The only parameter in the parton coalescence model is the width of Wigner function which can be related to the radius of the heavy meson or heavy (anti-)baryon~\cite{Song:2016lfv}, for example, according to the quark model \cite{Albertus:2003sx}.
Such width parameter is usually associated with a gaussian shape of the Wigner distribution which well describes the properties of the ground state of the hadron into which heavy quark coalesces. 

However, the problem is that the heavy quark can be projected into any physical states, for example, S-state,  P-state, D-state and so on.
Wigner functions for the excited states are more complicated than that of 
the ground state (S-state) and require additional parameters related to the physical radius of each state which is in most cases unknown.
Therefore most phenomenological models take into account the coalescence only into the ground state (and the first excited state) and rescale the coalescence probability with some normalization constant such that 
it reaches one in the limit $p\rightarrow 0$~\cite{Cao:2013ita}. 
Or the projection into all states available in the particle data book is carried out with a common radius for heavy mesons and another common radius for heavy baryons, which are arbitrarily taken such that the total probability reaches one at zero momentum and the ratio of $\Lambda_c$ to $D^0$ at intermediate momentum in heavy-ion collisions is reproduced~\cite{Oh:2009zj,Cho:2019lxb,Cao:2019iqs}.
In Ref.~\cite{Plumari:2017ntm}, when enhancing the heavy quark coalescence probability by using an amplified normalization factor of the Wigner density, the amplification/rescaling factor for baryons was taken to be squared of that for mesons, thereby also increasing the baryon-to-meson ratio by a rescaling factor.
Other models repeat the coalescence process in time until the accumulated coalescence probability is large enough at low momentum~\cite{Song:2015sfa,Song:2015ykw}, contrary to the assumption of the instant projection in the coalescence model.

One possible way to break through this complexity is adopting the statistical model~\cite{Andronic:2003zv}, which assumes that all hadron yields follow thermal equilibrium at a certain temperature and chemical potentials in the grandcanonical ensemble, which are called the chemical freeze-out temperature and chemical potential, respectively. 
By this assumption, the hadronization into the ground state of heavy flavor is treated by the parton coalescence model, while the hadronization into the excites states and even to heavy (anti-)baryons is realized simply by rescaling their coalescence probabilities according to the particle yield ratios from the statistical model without 
being bothered by their complicated Wigner functions and coalescence radii.
%\blue{However, although this model provides a solution to include the excited states into the hadronization process, it seems not be reliable for the case of heavy baryons at low transverse momentum. This is because the  baryon-meson ratio, which for charmed hadrons is $\Lambda_c/D^0$ and from the measurements by STAR and ALICE collaborations \cite{Adam:2019hpq,Acharya:2018ckj} is of the order of one, indicates an enhancement of the heavy baryon production which cannot be described by the thermal model. On the other hand, a coalescence model which is extended also to allow recombination of a heavy quark with two light quarks and form a heavy baryon \cite{Plumari:2017ntm} can explain such magnitude of the baryon-meson ratios, but as mentioned still suffers for the rescaling of coalescence probability.}
%\EB{Recently the enhancement of the ratio $\Lambda_c/D^0$ in heavy-ion collisions has been measured by the STAR and ALICE collaborations~\cite{Adam:2019hpq,Acharya:2018ckj}. It does mean that the statistical model does not work for charm-baryons, because the enhanced ratio is comparable to the baryon-to-meson ratios for light and strange hadrons and the yields of light and strange baryons are described by the statistical model~\cite{Andronic:2012dm}.}

In this paper we combine the coalescence model for the ground state of heavy meson and the statistical model for the excited states and (anti-)baryons in order to find out the conditions which overcome the insufficient coalescence probability at low momentum.
The main idea is to explore what kind of properties of partonic matter are more effective to lead 
a heavy quark in thermal equilibrium to hadronize at $T_c$ exclusively through coalescence.
As the first step of this approach we investigate the total coalescence probability of heavy quark at $T_c$.
We then speculate on what kind of nuclear matter enables the total coalescence probability of heavy quark to reach one, keeping the energy density 
at $T_c$ from the lattice QCD calculations~\cite{Borsanyi:2010cj,Borsanyi:2012cr}.
Considering that in the coalescence model the transition from unbound partons to hadrons instantly takes place, 
this study will provide a snapshot of a Quark-Gluon Plasma (QGP) just before hadronization through the parton coalescence.

The outline of the paper is as follows: In Sec.~\ref{coal-stat} we investigate total coalescence probablity of charm quark at $T_c$ by using the parton coalescence model and the statistical model for various parton masses and coalescence radii.
After that a search is made for the nuclear matter which realizes large enough coalescence probability of heavy quark, accompanied by its physical 
meaning in Sec.~\ref{modifications}.
The same approach is applied to bottom quark in Sec.~\ref{beauty} and summary is presented in Sec.~\ref{summary}.

\section{coalescence model+statistical model}\label{coal-stat}

Lte us briefly review the coalescence model where the number of produced particle through coalescence ($1+2\rightarrow 3$) is given by~\cite{Song:2016lfv}

\begin{eqnarray}
N_3=\frac{1}{(2\pi)^6}\frac{D_3}{D_1D_2}\int d^3{\bf r_1}d^3{\bf r_2}d^3{\bf p_1}d^3{\bf p_2}\nonumber\\
\times f_1({\bf r_1,p_1})f_2({\bf r_2,p_2})\Phi(r,k),
\label{coal}
\end{eqnarray}
where $D_i$ is the color-spin degeneracy factor of particle $i$ and the Wigner function of the formed hadron is given by the following gaussian distribution 
\begin{eqnarray}
\Phi(r,k)=8\exp\bigg[-\frac{r^2}{\sigma^2}-\sigma^2k^2\bigg]
\label{wignerF}
\end{eqnarray}
for the $1S$ state as a function of $r=|{\bf r_1-r_2}|$ %, ${\bf P=p_1+p_2}$, 
and $k=|m_2{\bf p_1}-m_1{\bf p_2}|/(m_1+m_2)$ in center-of-mass frame (${\bf p_1+p_2}=0$) with $\sigma$ being the width parameter in coordinate space.
Supposing
\begin{eqnarray}
f(p_1)=(2\pi)^3\delta^{(3)}({\bf p_1-p_c})\delta^{(3)}({\bf r_1-r_c}),
\end{eqnarray}
which means there is only one charm quark with three-momentum ${\bf p_c}=(p_c,0,0)$ in phase space, Eq.~(\ref{coal}) then turns to
\begin{eqnarray}
P(p_c)=\frac{1}{(2\pi)^3}\frac{D_3}{2N_c D_2}\int d^3{\bf r_2}d^3{\bf p_2}f_2(p_2)\Phi(r,k),
\label{coal-c}
\end{eqnarray}
where $D_1=2N_c$.
Eq.~(\ref{coal-c}) is the coalescence probability of  a single charm quark as a function of $p_c$ on which $k$ implicitly depends.
Spatial part is integrated out:
\begin{eqnarray}
\int d^3{\bf r_2}\exp\bigg[-\frac{r^2}{\sigma^2}\bigg]=\frac{1}{\gamma_{c.m.}}\int d^3{\bf r}\exp\bigg[-\frac{r^2}{\sigma^2}\bigg]\nonumber\\
=\frac{1}{\gamma_{c.m.}}(\sqrt{\pi}\sigma)^3,
\label{limit2}
\end{eqnarray}
where $\gamma_{c.m.}$ is the gamma factor of center-of-mass in heat bath frame.
The coalescence probability in Eq.~(\ref{coal-c}) is nonrelativistic. So it depends on frame and must be calculated in the center-of-mass frame by definition. Supposing a heavy quark has a large momentum, the center-of-mass frame of the heavy quark and a light anti-quark is quite different from the rest frame of a matter. As a result, anti-quark density in the center-of-mass frame is lower than the density in the rest frame of a matter, though this effect is small for low-momentum charm quark.
%
%Taking Galilean transformation (${\bf r=r_{c.m.}}$), for simplicity, which is valid for nonrelativistic $p_c$,
%\begin{eqnarray}
%\int d^3{\bf r_2}\exp\bigg[-\frac{r^2}{\sigma^2}\bigg]=(\sqrt{\pi}\sigma)^3,
%\label{limit1}
%\end{eqnarray}
%or assuming that center-of-mass frame is equivalent to the rest frame of 
%charm quark, which is valid in heavy-quark limit,
%\begin{eqnarray}
%\int d^3{\bf r_2}\exp\bigg[-\frac{r^2}{\sigma^2}\bigg]\approx\frac{1}{\gamma_c}\int d^3{\bf r}\exp\bigg[-\frac{r^2}{\sigma^2}\bigg]=\frac{1}{\gamma_c}(\sqrt{\pi}\sigma)^3,
%\label{limit2}
%\end{eqnarray}
%where the $\gamma_c$ is gamma factor of charm quark.
%
%Taking Eq.~(\ref{limit2}), the coalescence probability for $D$ meson ($D_3=1$) turns to
%
Finally the probability for charm quark coalescence to $1S$ state (or $D$ 
meson) is expressed as  
\begin{eqnarray}
P(p_c)=\frac{1}{2N_c}\bigg(\frac{\sigma}{\sqrt{\pi}}\bigg)^3 \int d^3{\bf p_2}\frac{e^{-\sigma^2k^2}}{\gamma_{c.m.}(e^{E_2/T}+1)}\nonumber\\
=\frac{\sigma^3}{N_c\sqrt{\pi}} \int dp_2p_2^2\int d\cos\theta\frac{e^{-\sigma^2k^2}}{\gamma_{c.m.}(e^{\sqrt{m_2^2+p_2^2}/T}+1)},
\label{integrals}
\end{eqnarray}
where $D_3=1$ and
\begin{eqnarray}
\frac{1}{D_2}f_2(p_2)=\frac{1}{e^{\sqrt{m_2^2+p_2^2}/T}+1}.
\end{eqnarray}
The width parameter $\sigma$ is related to the $D$ meson radius~\cite{Song:2016lfv}:
\begin{eqnarray}
\sigma^2=\frac{4}{3}\frac{(m_c+m_q)^2}{m_c^2+m_q^2}\langle r_D^2\rangle.
\end{eqnarray}

Taking into account the thermal motion of charm quark, thermal averaged coalescence probability of charm quark is given by
\begin{eqnarray}
\langle P(p_c)\rangle=\frac{1}{n_c}\int \frac{d^3 p_c}{(2\pi)^3} \frac{P(p_c)}{e^{\sqrt{m_c^2+p_c^2}/T}+1},
\label{thermal-average}
\end{eqnarray}
where
\begin{eqnarray}
n_c=\int \frac{d^3 p_c}{(2\pi)^3}  \frac{1}{e^{\sqrt{m_c^2+p_c^2}/T}+1}.
\end{eqnarray}

According to the lattice calculations~\cite{Borsanyi:2010cj,Borsanyi:2012cr}, energy density at the critical temperature for $\mu_B=0$ ($T_c\approx 
0.16~ {\rm GeV}$) is around ${\rm 0.4~ GeV/fm^3}$, which can be interpreted as 0.45 GeV of light quark mass in quasi-particle picture~\cite{Plumari:2011mk,Moreau:2019vhw} and 0.55 GeV of strange quark mass, taking into account the mass difference between $D$ and $D_s$.
We ignore gluon by assuming that it has already split into quark and antiquark at $T_c$ for coalescence.

\begin{figure}[th!]
\centerline{
\includegraphics[width=9 cm]{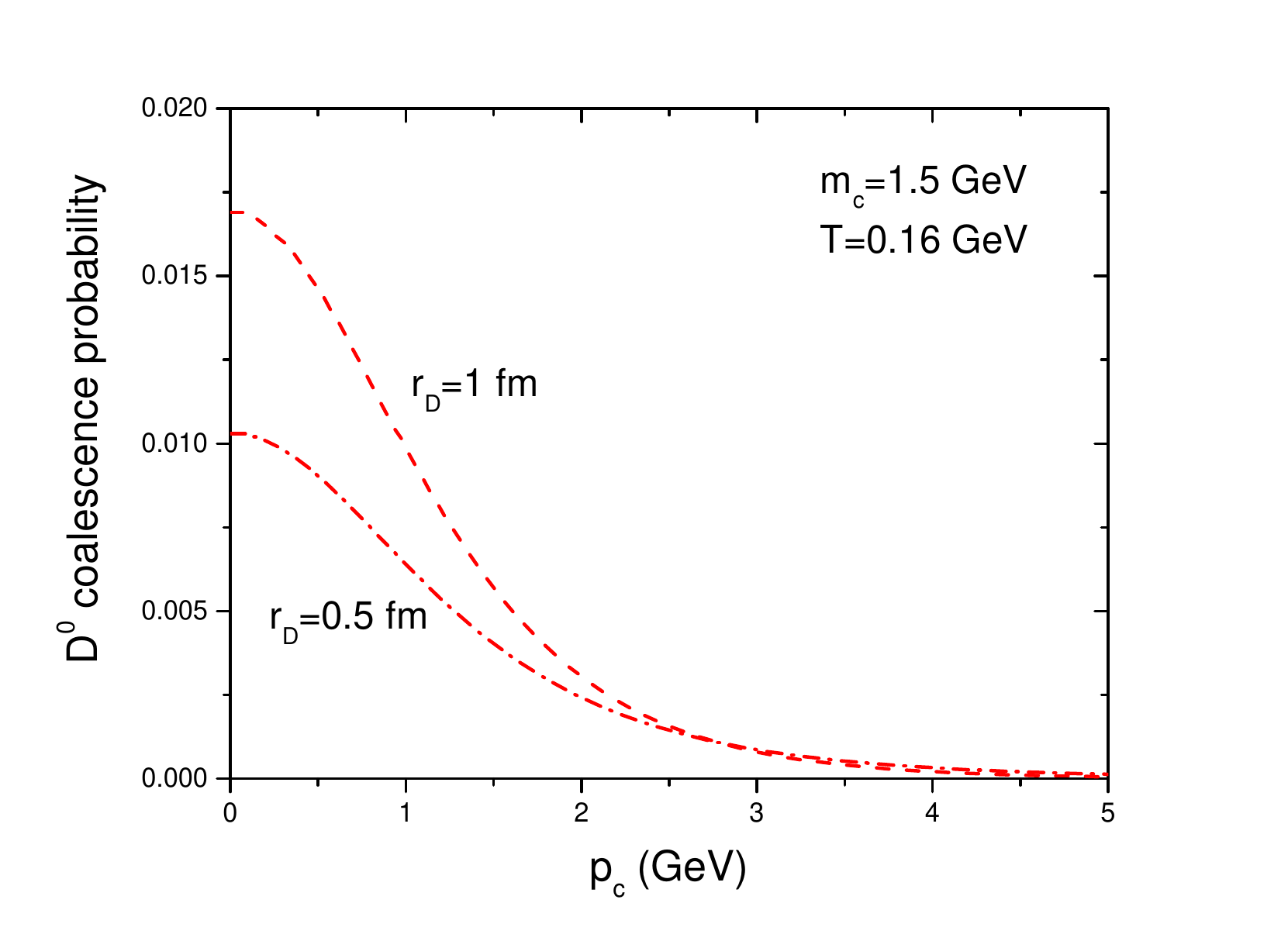}}
\caption{(Color online) coalescence probability of charm to form $D^0$($D^+$) at $T_c=$ 0.16 GeV for $r_D=$ 0.5 fm and 1 fm.
$m_c$ is taken to be 1.5 GeV}
\label{coal-prob}
\end{figure}

Figure~\ref{coal-prob} shows the coalescence probability of charm to form 
$D^0$ or $D^+$ at $T_c=$ 0.16 GeV for $r_D=$ 0.5 fm and 1 fm with $m_c$ taken to be 1.5 GeV.
The probability decreases with charm quark momentum.
From Eq.~(\ref{thermal-average}) the thermal averaged coalescence probability of charm quark is roughly 0.71 \% and 1.1 \%  for $r_D=$ 0.5 fm and 1 fm, respectively.

The same coalescence model can be used for the formation of excite states, for example, $D^*$.
However, we for simplicity assume that the statistical model works and find at $T_c$

\begin{eqnarray}
\frac{{\rm number~ density~ of~ all~ charm~ hadrons}}{{\rm number~ density~ of~} D^0}=7.47,
\label{ratio-charm}
\end{eqnarray}
considering all hadrons which contains charm flavor~\cite{pdg,Andronic:2021erx}.
Recently the enhancement of the ratio $\Lambda_c/D^0$ in heavy-ion collisions has been measured by the STAR and ALICE collaborations~\cite{Adam:2019hpq,Acharya:2018ckj}.
Roughly estimated, $\Lambda_c$ is twice enhanced compared to that from the statistical model, which can be explained by including the charm baryon states that have not observed in experiment but possibly exist according to the relativistic quark model~\cite{He:2019tik}.
Assuming that charm baryons are enhanced twice, the ratio in Eq.~(\ref{ratio-charm}) increases to 8.74, if the statistical model works and the unobserved charm baryon states really exist.
However, it does not change our results qualitatively.
From Eq.~(\ref{ratio-charm}) the total coalescence probability of charm quark is about 6.2 \% and 
9.8 \% for $r_D=$ 0.5 fm and 1 fm, respectively.

\begin{figure}[th!]
\centerline{
\includegraphics[width=9 cm]{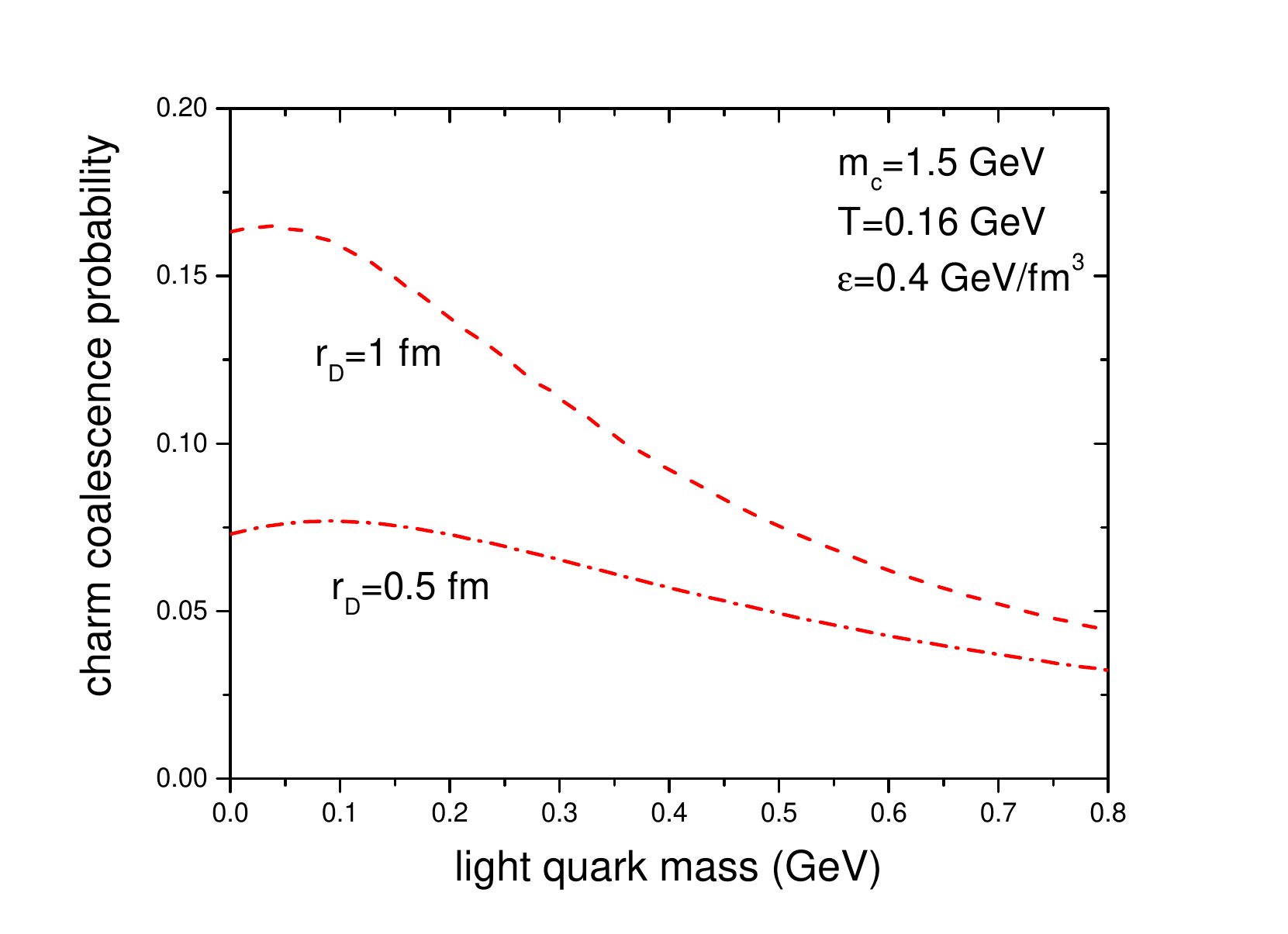}}
\caption{(Color online) total coalescence probability of charm at $T_c=$ 0.16 GeV as a function of light quark mass for the same energy density $\epsilon=0.4 ~{\rm GeV/fm^3}$. }
\label{coal-prob2}
\end{figure}

In order to see the effects of light (anti-)quark mass on the coalescence probability of charm quark, we vary light quark mass, keeping energy density and the relation $m_s-m_q=$ 0.1 GeV by rescaling light quark number 
density.

Figure~\ref{coal-prob2} shows the total coalescence probability of charm at $T_c=$ 0.16 GeV as a function of light quark mass.
One can see that the total coalescence probability increases with decreasing light quark mass. However, it is at most around 16-17 \% for massless 
quark and $r_D=$ 1 fm, which implies that the environments for charm quark coalescence cannot simply be thermal light (anti-)quarks which are completely randomized in phase space.

%For information we show in figure~\ref{number-density} number density of 
%antiquarks which are candidates for the coalescence of charm quark and interspace between antiquarks.

%\begin{figure}[th!]
%\centerline{
%\includegraphics[width=12 cm]{number-density.pdf}}
%\caption{(Color online) number density of antiquarks and interspace between them as a function of light quark mass at T=0.16 GeV and energy density of 0.66 ${\rm GeV/fm^3}$}
%\label{number-density}
%\end{figure}

In Refs.~\cite{Oh:2009zj,Cho:2019lxb,Cao:2019iqs} the total coalescence probability of charm quark reaches one at zero momentum by including all available states. 
However, they are different from our approach for a couple of reasons.
Firstly, quark mass there is taken to be 300 MeV which is lighter than 450 MeV that reproduces the energy density at the critical temperature from lattice QCD~\cite{Plumari:2011mk,Moreau:2019vhw}.
From the next section quark mass will be fixed at 450 MeV. 
Secondly, the critical temperature in Ref.~\cite{Oh:2009zj,Cho:2019lxb} is taken to be 175 MeV, which is higher than 160 MeV of the present study.
Both differences increase the number density of (anti)quark at $T_c$ and, as a result, the total coalescence probability of charm quark enhances about twice.
The most important difference is that they use a common coalescence radius for all charm mesons.
In this case the coalescence probability for $D^*$ is simply three times larger than that for $D$ meson, since energy is not taken into account in Eq.~(\ref{wignerF}).
On the other hand, the statistical model predicts about 1.37 times larger $D^*$ number density than $D$ number density at $T=160$ GeV because of the mass difference between $D^*$ and $D$ mesons.
Normally the number density of excited state is enhanced by the spin degeneracy but suppressed by the larger mass in the statistical model.
However, the coalescence model with a common radius neglects the latter effect and, as a result, exaggerates the number density of excited states, compared to the statistical model.   
We note that space-momentum correlations of heavy quarks with partons may enhance coalescence probability~\cite{He:2019vgs}.

\section{modifications of nuclear matter at $T_c$}\label{modifications}

In heavy-ion collisions heavy quarks hadronize through coalescence at low 
momentum while they mostly undergo fragmentation at high momentum~\cite{He:2014cla,Cao:2013ita,Gossiaux:2010yx,Das:2015ana,Song:2015sfa,Song:2015ykw,Plumari:2017ntm}.
However, as shown in the previous section, the total coalescence probability of charm quark does not reach one even at low momentum.
In this section we look for the environments where the coalescence probability is close to one, which, as we inferred in previous section, must be different from randomized thermal distribution of (anti-)light quarks.

\begin{figure}
  \centering
  \begin{tikzpicture}
    \begin{feynman}
      \vertex(q1i) at (-3, 2) {$q$};
      \vertex(a1i) at (-1, 2) {$\bar{q}$};
      \vertex(q2i) at ( 1, 2) {$q$};
      \vertex(a2i) at ( 3, 2) {$\bar{q}$};%{$\uparrow$};
      \vertex(q1f) at (-2.4, -2) {$q$};
      \vertex(a1f) at (-1.6, -2) {$\bar{q}$};
      \vertex(q2f) at ( 1.6, -2) {$q$};
      \vertex(a2f) at ( 2.4, -2) {$\bar{q}$};%{$\uparrow$};
      \vertex(q1m) at (-2.7, 0);
      \vertex(a1m) at (-1.3, 0);
      \vertex(q2m) at ( 1.3, 0);
      \vertex(a2m) at ( 2.7, 0);%{$\uparrow$};
      \vertex(ex1) at (-1.2, 0.1);
      \vertex(ex2) at (1.2, 0.1);
      \vertex(ran1) at (-3.5, 1.5);
      \vertex(ran2) at (4.2, 1.5){$randomized$};
      \vertex(cor1) at (-3.5, -1.5);
      \vertex(cor2) at (4.2, -1.5){$correlated$};
      \diagram* {
        (q1i) -- [fermion, out=-90, in=135] (q1m) -- [fermion, out=-45, in=90] (q1f),
        (a1f) -- [fermion, out=90, in=-135] (a1m) -- [fermion, out=45, in=-90] (a1i),
        (q2i) -- [fermion, out=-90, in=135] (q2m) -- [fermion, out=-45, in=90] (q2f),
        (a2f) -- [fermion, out=90, in=-135] (a2m) -- [fermion, out=45, in=-90] (a2i),
        (q1m) -- [gluon] (a1m), 
        (q2m) -- [gluon] (a2m), 
        (ex1) -- [gluon] (ex2),
        (ran1) -- [scalar] (ran2),
        (cor1) -- [scalar] (cor2),
      };
    \end{feynman}
  \end{tikzpicture}
  \caption{schematic diagram of the transition from randomized phase to correlated phase. There must be gluon exchange between pairs to conserve energy in the coalescence model. The coalescence model ought to be taken at the correlated phase, not at the randomized phase}
\label{diagram}
\end{figure}
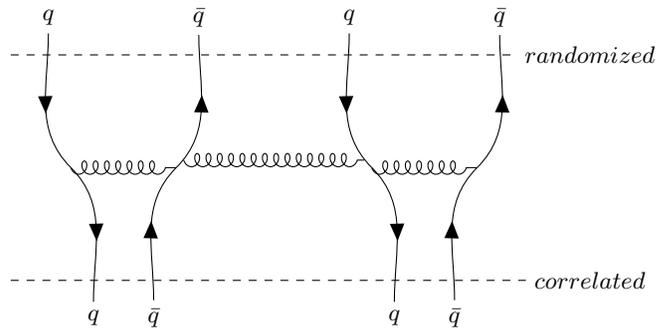

The phase transition from QGP to hadron gas, in other words, the hadronization of heavy quark is known as a crossover at low baryon chemical potential, which means that heavy quark smoothly hadronizes as shown in figure~\ref{diagram}. The most significant difference between QGP and hadron gas is the correlations between quark and antiquark.
In QGP quark and antiquark are completely uncorrelated as in the upper part of figure~\ref{diagram}, while in hadron gas quark and antiquark are strongly correlated with each other in color, momentum and space as in the lower part of figure~\ref{diagram}. Our aim is to find when the instant coalescence formalism should be applied for all the heavy quarks at low momentum to be consumed by the coalescence. For this the correlations between quark and antiquark in color, momentum and spatial spaces are necessary. Naively thinking, color correlation must first take place for quark-antiquark pair to be a color singlet by emitting or absorbing soft gluons, and then momentum correlation will take place due to the attractive force of the color singlet, which will be followed by the spatial correlation, because quark and antiquark will be closer to each other.

\subsection{color rearrangement}
One may assume that color charges around the charm quark are not randomly 
distributed but rearranged for the charm to easily form a color singlet, just as in the breaking of a color string.
We assume that the probability for an antiquark to have the opposite color of charm quark such that they form a color singlet is a function of spatial distance between them as following:

\begin{eqnarray}
\frac{N_c-1}{N_c}\exp\bigg[-\frac{r^2}{\sigma_c^2}\bigg]+\frac{1}{N_c},
\label{lengthc}
\end{eqnarray}
where $\sigma_c$ is the color correlation length, such that the probability to be the opposite color is one at $r=0$ and $1/N_c$ at large distance. 
Eq.~(\ref{lengthc}) enhances the total color charge opposite to that of charm quark.
In principle, the anti-color charge should be reduced at large distance to compensate the local condensation near charm quark. 
However, we ignore it for simplicity, because coalescence probability exponentially decreases with distance and the depletion at large distance will barely affects the total coalescence probability.

Eq.~(\ref{limit2}) is then modified into

\begin{eqnarray}
\int d^3{\bf r_2}\exp\bigg[-\frac{r^2}{\sigma^2}\bigg]\rightarrow~~~~~~~~~~~~~~~~~~~~~~~~\nonumber\\
 \frac{1}{\gamma_{c.m.}}\int d^3{\bf r}\bigg\{(N_c-1)\exp\bigg[-\frac{(\sigma^2+\sigma_c^2)r^2}{\sigma^2\sigma_c^2}\bigg]\nonumber\\
+\exp\bigg[-\frac{r^2}{\sigma^2}\bigg]\bigg\}\nonumber\\
=\frac{\pi^{3/2}}{\gamma_{c.m.}}\bigg\{(N_c-1)\bigg(\frac{\sigma^2\sigma_c^2}{\sigma^2+\sigma_c^2}\bigg)^{3/2}+\sigma^3\bigg\}.
\label{color-corr}
\end{eqnarray}

%The above modification will slightly enhance the total canti-red color charge of the matter.
%Assuming an infinite matter, the small change will be negligable.

\begin{figure}[t!]
\centerline{
\includegraphics[width=9 cm]{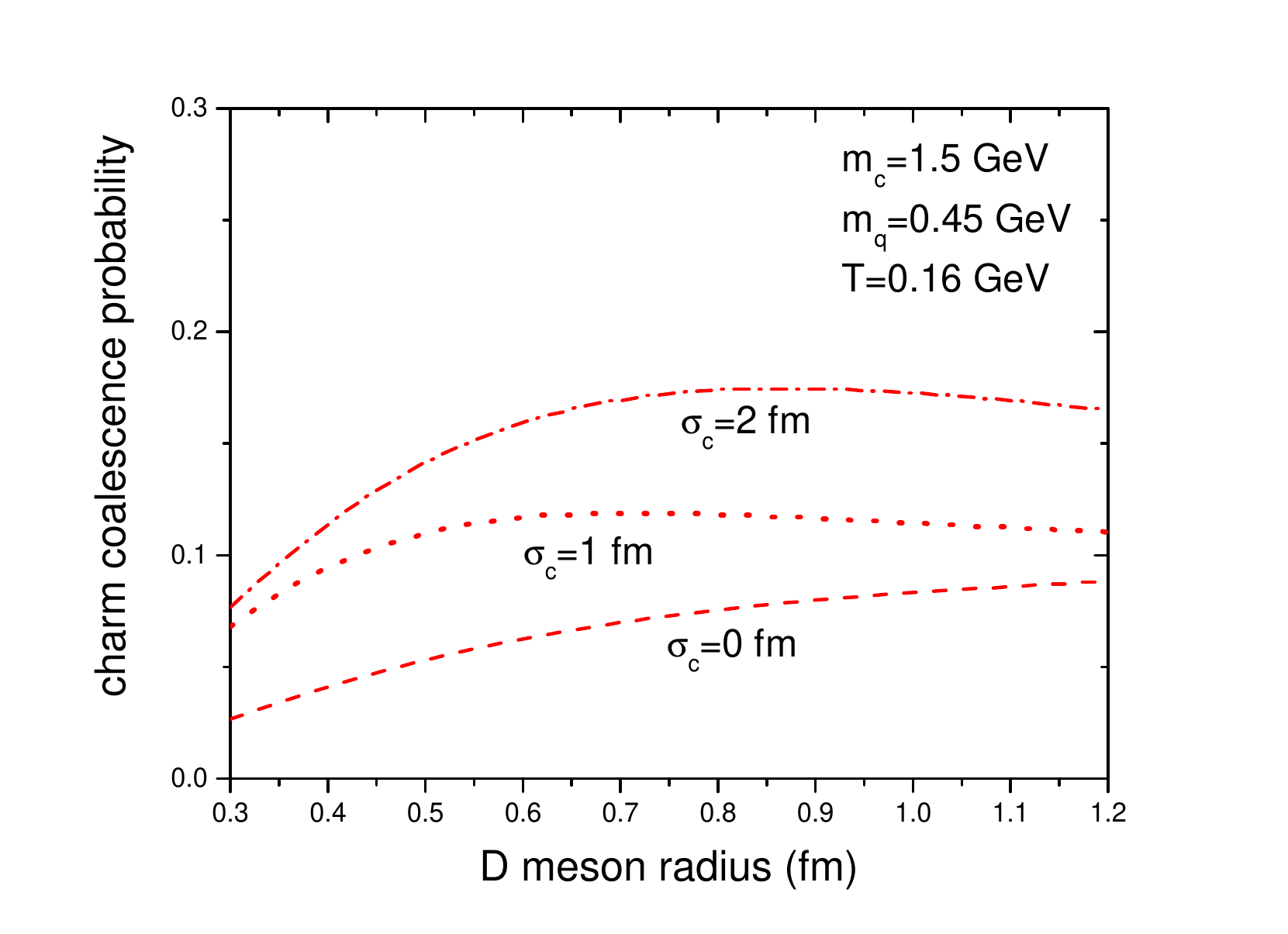}}
\caption{(Color online) total coalescence probability of charm as a function of $D$ meson radius for a couple of color correlation strength.}
\label{coal-prob3}
\end{figure}

Figure~\ref{coal-prob3} shows the total coalescence probability of charm as a function of $D$ meson radius for the light (anti-)quark mass $m_q=$ 
0.45 GeV.
The color correlation enhances the total coalescence probability, though the probabilities are still far below one.
In the limit of vanishing correlation length ($\sigma_c\rightarrow 0$) Eq.~(\ref{color-corr}) returns to Eq.~(\ref{limit2}), while for infinite correlation length ($\sigma_c\rightarrow \infty$) Eq.~(\ref{color-corr}) converges to

\begin{eqnarray}
\int d^3{\bf r_2}\exp\bigg[-\frac{r^2}{\sigma^2}\bigg]\rightarrow \frac{N_c}{\gamma_{c.m.}}(\sqrt{\pi}\sigma)^3,
\end{eqnarray}
which is nothing but the coalescence probability without color factor, that is, three times the dashed line in Figure~\ref{coal-prob3}.
It is the maximum coalescence probability of charm quark which can be obtained by the color correlations.
Therefore, we can conclude that the color correlation is not enough to describe the environments for charm quark coalescence at $T_c$.

\subsection{momentum and/or spatial rearrangement}

Now we try momentum rearrangement by introducing a deformation to the light (anti-)quark distribution $f_2(p_2)$:
\begin{eqnarray}
f_2(p_2)\rightarrow N_2(p_c,\Delta) D_2 \frac{e^{-\Delta^2 k^2}}{e^{E_2/T}+1},
\label{p-correlation}
\end{eqnarray}
where $D_2$ is spin-color degeneracy factor, $k$ is relative momentum in center-of-mass frame as in Eq.~(\ref{coal}) and  $N_2(p_c,\Delta)$ is the 
normalization factor to keep the number density of light (anti-)quarks unchanged, that is,

\begin{eqnarray}
N_2(p_c,\Delta) \equiv ~~~~~~~~~~~~~~~~~~~~~~~~~~~~~~~~~~~~~~~~~~~~\nonumber\\
\int d^3{\bf p_2}\frac{1}{\gamma_{c.m.}(e^{E_2/T}+1)} \bigg/ \int 
d^3{\bf p_2}\frac{e^{-\Delta^2 k^2}}{\gamma_{c.m.}(e^{E_2/T}+1)}.
\end{eqnarray}
$\Delta$ in Eq.~(\ref{p-correlation}) indicates the strength of momentum correlations. Infinite $\Delta$ stands for the strongest correlation, that is, ${\bf p_2}={(m_q/m_c)\bf p_c}$, while vanishing $\Delta$ means no momentum
correlation and light (anti-)quark distribution returns to normal Fermi-Dirac distribution:
\begin{eqnarray}
f_2(p_2)= D_2 \frac{1}{e^{E_2/T}+1}.
\end{eqnarray}

It is reasonable to assume that the correlation strength depends on the distance from charm quark, for example,
\begin{eqnarray}
%\Delta= \tan^{-1}(r/r_D),
\Delta= \Delta_0 e^{-d/\sigma_c},
\label{delta}
\end{eqnarray}
where $d$ is the distance from charm quark in charm rest frame and $\Delta_0$ is the momentum corrleation at $d=0$.
We may use the same correlation length as the color correlation length $\sigma_c$, because the momentum correlation is induced by color interactions.

Substituting Eqs.~(\ref{p-correlation}) and (\ref{delta}) into Eq.~(\ref{coal-c}), coalescence probability for $D$ meson is modified into
\begin{eqnarray}
P(p_c)=\frac{1}{(2\pi)^3}\frac{1}{2N_c}\int d^3{\bf r_2}\int d^3{\bf p_2}\nonumber\\
\times N_2(p_c,\Delta)\frac{e^{-\Delta^2 k^2}}{e^{E_2/T}+1}\Phi(r,k)\nonumber\\
=\frac{1}{2N_c\pi^3}\int d^3{\bf r_2}e^{-r^2/\sigma^2}N_2(p_c,\Delta)\nonumber\\
\times \int d^3{\bf p_2}\frac{e^{-(\Delta^2+\sigma^2) k^2}}{e^{E_2/T}+1}.
\end{eqnarray}

In heavy quark limit where center of mass frame is equivalent to charm rest frame ($d=r$), 
\begin{eqnarray}
P(p_c)\approx\frac{1}{2N_c\pi^3}\int d^3{\bf r}e^{-r^2/\sigma^2}N_2(p_c,\Delta)\nonumber\\
\times \int d^3{\bf p_2}\frac{e^{-(\Delta(r)^2+\sigma^2) k^2}}{e^{E_2/T}+1}\frac{1}{\gamma_{c.m.}}\nonumber\\
=\frac{4}{N_c\pi}\int dr r^2 e^{-r^2/\sigma^2}N_2(p_c,\Delta)\nonumber\\
\times\int dp_2 p_2^2\int d\cos\theta \frac{e^{-(\Delta(r)^2+\sigma^2) k^2}}{e^{\sqrt{m_2^2+p_2^2}/T}+1}\frac{1}{\gamma_{c.m.}},
\label{p-corr}
\end{eqnarray}
where $\Delta(r)= \Delta_0 e^{-r/\sigma_c}$.
One can see that Eq.~(\ref{p-corr}) turns to Eq.~(\ref{integrals}) in the 
limit $\Delta_0\rightarrow 0$.

\begin{figure}[t!]
\centerline{
\includegraphics[width=9 cm]{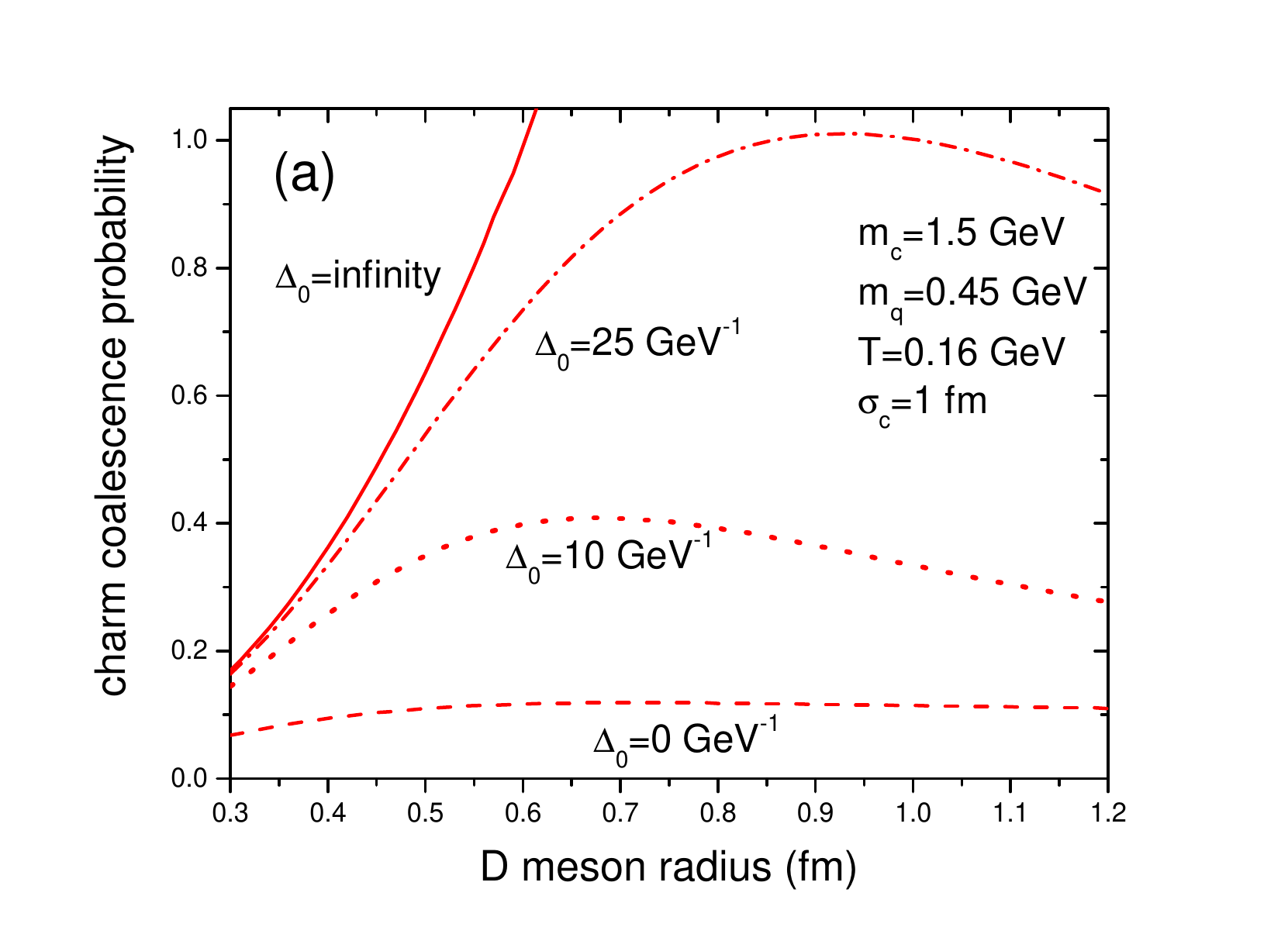}}
\centerline{
\includegraphics[width=9 cm]{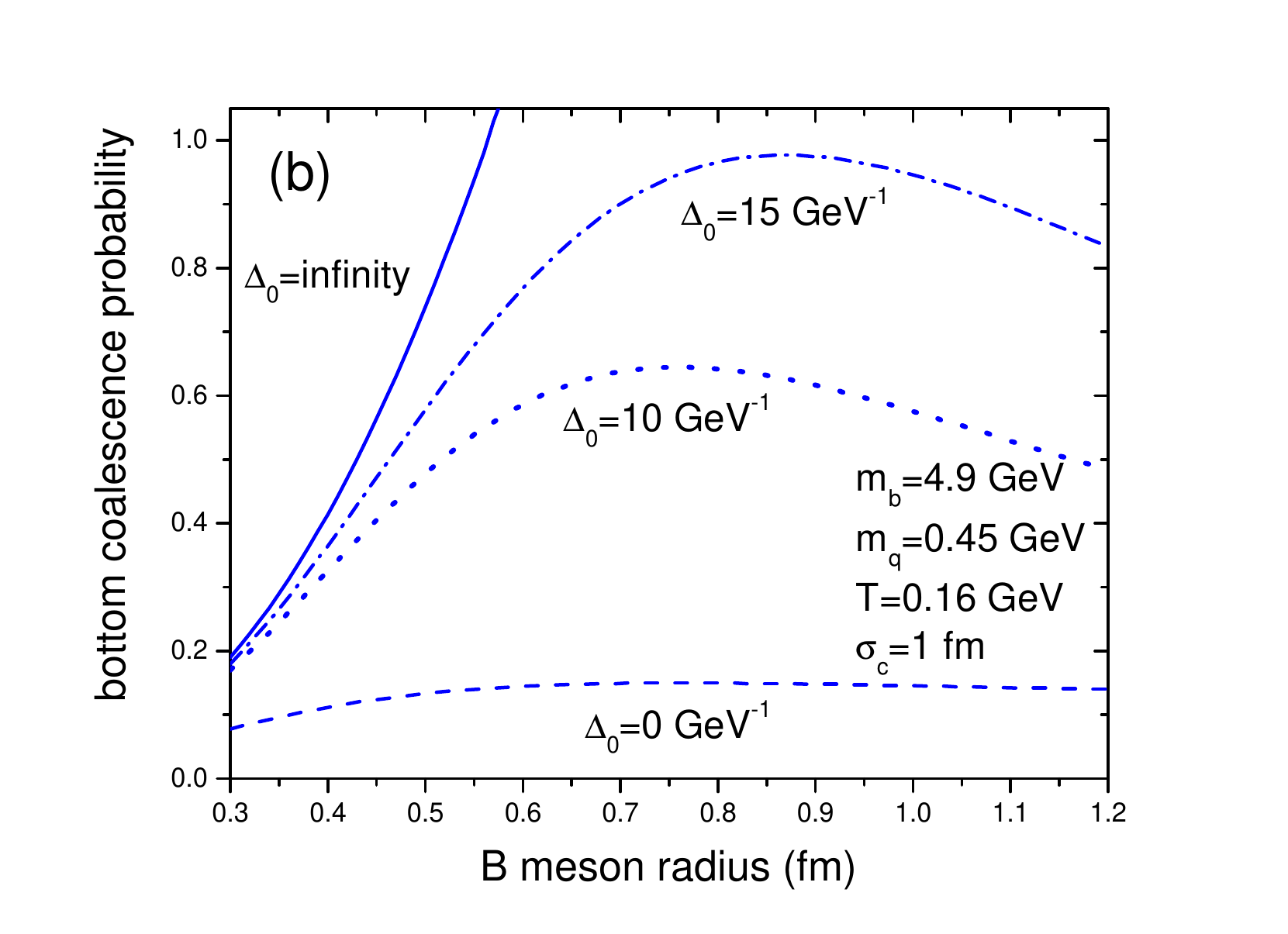}}
\caption{(Color online) total coalescence probability of charm(bottom) as 
a function of $D$($B$) meson radius for several momentum correlation strengths with color and momentum correlation length being 1 fm in upper(lower) panels}
\label{coal-probp}
\end{figure}

We show in figure~\ref{coal-probp} (a) the total coalescence probability of charm quark with color and momentum correlations as a function of $D$ meson radius for several momentum correlation strength with the color and 
momentum correlation length being 1 fm.
One can see that the effects of momentum correlation on coalescence probability are stronger than those of color correlation and total coalescence 
probability reaches one for a strong momentum correlation, depending on $D$ meson radius. 

For a small radius of $D$ meson the coalescence probability is less sensitive to the relative momentum between charm and light antiquark, which reduces the effects of momentum correlation on coalescence probability.
If $D$ meson radius is less than 0.6 fm at $T_c$, another correlation such as spatial correlation will be needed for the coalescence probability to reach one.
It can be realized by introducing $r-$dependent $N_2$, that is, $N_2(p_c,\Delta,r)$.
However, the spatial correlation will take place last, because it is induced by the correlation of color and momentum.

We note that the total coalescence probability increases above 1 for a large radius in figure~\ref{coal-probp}, which looks unphysical. However, it is an artifact from the assumption that the coalescence probability of charm quark with one single parton is not affected by the presence of other partons in the medium.
Considering that the charm quark cannot make coalescence with two partons simultaneously to form a $D$ meson, the coalescence probability with one parton is not completely independent from that computed for the other parton~\cite{Lin:2009tk}.
This overlapping problem emerges in dense medium or when the contributions of different partons to the charm coalescence probability are not negligible compared to one.
Therefore, the solid line in figure~\ref{coal-probp} will, in reality, smoothly converges to 1 with increasing $D$ meson radius. 

We can speculate that the presence of strong momentum correlations is physically motivated by the strong non-perturbative nature of heavy quark interactions with the QGP around $T_c$. This has been investigated in different studies, where, for example, the temperature dependence of heavy quark transport coefficients within non-perturbative quasi-particle 
approaches is shown to provide correct description of various heavy quark 
observables~\cite{Song:2015sfa,Scardina:2017ipo} and also to give an estimate of spatial diffusion coefficients in agreement with lQCD calculations~\cite{Banerjee:2011ra}. However, a precise derivation of the origin of such momentum correlations and their relation to the properties of heavy quark interactions with the QGP can be done only by coupling our hadronization model to a dynamical evolution of QGP matter near $T_c$, which is far from the setup of this study. We remind here that the goal of this work is to provide a snapshot of the QGP properties at hadronization through coalescence of heavy quark at $T_c$.

\section{extension to bottom}\label{beauty}

The same procedures can be repeated for bottom quark.
Bottom quark mass is taken to be 4.9 GeV from the mass difference between 
$D$ and $B$ mesons.
From the statistical model at T= 0.16 GeV,

\begin{eqnarray}
\frac{{\rm number~ density~ of~ all~ bottom~ hadrons}}{{\rm number~ density~ of~} B^-}=11.4.
\end{eqnarray}

Figure~\ref{coal-probp} (b) shows the coalescence probability of bottom quark with color and momentum correlations as a function of $B$ meson radius.
The coalescence probability for $\Delta_0=0~ {\rm GeV^{-1}}$ corresponds to the probability only with color correlation as in Figure~\ref{coal-probp} (a).
Similar to charm, the coalescence probability can reach one by a strong momentum correlation.
Though it seems that less strong momentum correlation is needed for bottom quark, compared to charm quark, one should keep it in mind that the radius of $B$ meson is smaller than that of $D$ meson. 
Since bottom quark is much heavier than charm quark, bottom is more static in heat bath, which makes the coalescence probability of bottom shown in Figure~\ref{coal-probp} (b) always larger than that of charm in Figure~\ref{coal-probp} (a) for the same radius and strength of color/momentum correlations.

\section{summary}\label{summary}

We have calculated the total coalescence probability of heavy quark at $T_c$ by using the parton coalescence model for the formation of the ground 
state and then the statistical model for the formation of excited states as well as (anti-)baryons.
It has been found that the total coalescence probability of heavy quark is far below one in a quark-gluon plasma composed of thermal quarks and antiquarks which are completely randomized in color, momentum and spatial spaces.
Assuming that all heavy quarks which have only thermal momentum at $T_c$ 
hadronize through parton coalescence, the correlations of heavy quark and 
light antiquark are inevitable before hadronization.
We have found that the correlation of color charges is not enough  for a sufficient coalescence probability but the correlation in momentum space is also required.
For a small radius of heavy meson at $T_c$ the correlation in coordinate space is additionally needed. 

The coalescence model does not conserve energy and the phase transition cannot take place instantly~\cite{Song:2016lfv}.
In fact the phase transition at small baryon chemical potential is known as crossover according to the lattice calculations~\cite{Aoki:2006we}.
More realistic picture will be a gradual hadronization as shown in figure~\ref{diagram}.
For example, quarks and antiquarks are rearranged in color, momentum and coordinate spaces with time by emitting and absorbing soft gluons and then the coalescence model will be applicable, because energy conservation will be satisfied at this moment.
As in figure~\ref{diagram}, gluon exchange must take place not only between quark and antiquark pair which will form a bound state but also with quark or antiquark of other pairs, which solves the problem of energy conservation in the coalescence model.
The coalescence model is normally applied to the randomized phase which is indicated by the upper dashed line in figure~\ref{diagram}, but we argue that it must be applied to the correlated phase which is shown by the lower dashed line. 
The purpose of our study is not to describe how quarks and antiquarks are rearranged for hadronization but to provide a snapshot before hadronization, and our finding is
 that partons close to hadronization, that is, when the coalescence model is applicable, are not completely randomized in phase space but strongly correlated in color, momentum and/or coordinate spaces, ready to form bound states.

\section*{Acknowledgements}
The authors acknowledge valuable discussions with Elena Bratkovskaya and her suggestions.
We thank CRC-TR 211 'Strong-interaction matter under extreme conditions'- 
project Nr.315477589 - TRR 211.
The computational resources have been provided by the LOEWE-Center for Scientific Computing and the "Green Cube" at GSI, Darmstadt.

\end{document}